\documentclass[twocolumn,showpacs,prb,10pt,aps,floatfix]{revtex4-1}
\usepackage[english]{babel}
\usepackage{amsmath,amssymb}
\usepackage{times}
\usepackage{epsfig}
\usepackage[colorlinks,linkcolor=blue,citecolor=blue,urlcolor=blue]{hyperref}
\usepackage{color}
\begin{document}
\title{Antiferromagnetic order in weakly coupled random spin chains}
\author{J. Kokalj$^{1}$}
\author{J. Herbrych$^{2}$}
\author{A. Zheludev$^{3}$}
\author{P. Prelov\v{s}ek$^{1,4}$}
\affiliation{$^1$J. Stefan Institute, SI-1000 Ljubljana, Slovenia}
\affiliation{$^2$Crete Center for Quantum Complexity and Nanotechnology,
Department of Physics, University of Crete, P.O. Box 2208, 71003
Heraklion, Greece} 
\affiliation{$^3$
Neutron Scattering and Magnetism, Laboratory for Solid State Physics,
ETH Z\"urich, Z\"urich, Switzerland}
\affiliation{$^4$Faculty of Mathematics and Physics, University of Ljubljana,
SI-1000 Ljubljana, Slovenia}
\date{\today}

\begin{abstract}
The ordering of weakly coupled random antiferromagnetic $S=1/2$
chains, as relevant for recent experimentally investigated spin
chain materials, is considered theoretically. The one-dimensional
isotropic Heisenberg model with random exchange interactions is
treated numerically on finite chains with the density-matrix
renormalization-group approach as well as with the standard
renormalization analysis, both within the mean-field approximation for 
interchain coupling $J_{\perp}$. Results for the ordering temperature 
$T_N$ and for the ordered moment
$m_0$ are presented and are both reduced with the increasing
disorder agreeing with experimental observations. 
The most pronounced effect of the random singlet concept appears to be
a very large span of local ordered moments, becoming wider with decreasing
$J_{\perp}$, consistent with $\mu$SR experimental findings.
\end{abstract}
\maketitle

\section{Introduction}
The antiferromagnetic (AFM) Heisenberg model of $S=1/2$ spins on a
one-dimensional (1D) chain represents one of the prototype and most
studied quantum many-body model for strongly correlated electrons,
being at the same time realized nearly perfectly in several
materials. Since 1D spin systems do not exhibit any long range order
even at temperature $T=0$, the ordering appears through the interchain
coupling. The ordering N\'eel temperature $T_N$ emerging in weakly
coupled AFM chains is by now well described theoretically
\cite{schulz96}, being confirmed by numerical calculations
\cite{yasuda05} and experimental investigations on materials with
quasi-1D spin systems \cite{tsukada99}.

The quenched disorder in intrachain exchange couplings reveals in 1D
spin chains qualitatively new phenomena as well theoretical and
experimental challenges. Even in the case of unfrustrated AFM random
Heisenberg chain (RHC) it has been shown using the
renormalization-group (RG) approaches
\cite{ma79,dasgupta80,fisher94,westerberg95} that the $T \to 0$
behavior is qualitatively changed by any disorder leading to the
concept of random singlets (RS). The signature of such state is the
singular - Curie-like - divergence of the uniform susceptibility
$\chi_0(T \to 0)$ \cite{hirsch80}.  Such behavior was
also found for exactly solvable model of impurities coupled with
random exchange interactions to the host Heisenberg chain, but only
for strong randomness \cite{klumper98a}.  Refreshed theoretical
interest in RHC phenomena has been stimulated by the synthesis and
experimental investigations of novel materials representing the
realization of RHC, in particular BaCu$_2$(Si$_{1-x}$Ge$_x$)$_2$O$_7$
\cite{yamada01,tsukada99,shiroka11} and
Cu(py)$_2$(Cl$_{1-x}$Br$_x$)$_2$ \cite{thede12} compounds.
Experiments confirmed theoretically predicted $\chi_0(T)$
\cite{zheludev07}, but revealed also novel features as large and
strongly $T$-dependent spread of local NMR spin-lattice relaxation
times \cite{shiroka11,shiroka13} which has been reproduced within the
simple RHC model \cite{herbrych13}.

The existence of weak but finite interchain couplings $J_{\perp}$ in
quasi-1D RHC compounds and related AFM ordering at low $T<T_N$ open a
new perspective on the RS systems \cite{thede12}. Mixed
BaCu$_2$(Si$_{1-x}$Ge$_x$)$_2$O$_7$ \cite{yamada01} as well
Cu(py)$_2$(Cl$_{1-x}$Br$_x$)$_2$ \cite{thede12} show a substantial
reduction of $T_N$ as well as the ground state (g.s.) $T=0$ ordered magnetic 
moment $m_0$ relative to the disorder-free ($x=0, 1$) materials. 
Theoretical treatments so far suggested even the opposite trend
\cite{joshi03} 
revealing the difficulties of theoretical approaches. The central
theoretical issue also in connection with experiments is to what
extent and in which properties the singular behavior of quantum RS
physics remains reflected in the long-range AFM order at low $T$. 
The aim of this paper is to present results of numerical and
analytical calculations which show that under the presence of
weak (but not extremely weak) interchain coupling 
treated within a mean-field approximation (MFA)
randomness reduces both $T_N$ as well as $m_0$, which is in agreement
with experiment. We also present evidence
that the RS phenomena is reflected in large distribution of $T=0$
local ordered moments $m_{i}$ being consistent with preliminary
experimental results \cite{thede14}. 

The paper is organised as follows. In Section \ref{sec:mf} we
introduce the model and the MFA approximation. In Section \ref{sec:ssnt} we
introduce the numerical method and present results on staggered
susceptibility and transition temperature. This is followed by
presentation of results for ordered moments and their distribution in
Section \ref{sec:om}. In Section \ref{sec:rg} we discuss results
obtained by RG and in Section \ref{sec:exp} we compare our results in
more detail with experiment. Conclusions are given at the end in 
Section \ref{sec:diss}.

\section{Model}
\label{sec:mf}
Our goal is to understand properties in particular the ordering in
the quasi-1D RHC model, which is given by quenched (intrachain)
random exchange couplings $J_{i,j}$ and constant interchain coupling $J_{\perp}$,
\begin{equation}
H= \sum_{i,j} J_{i,j}\,\mathbf{S}_{i,j}\cdot\mathbf{S}_{i+1,j}
+J_{\perp} \sum_{i,\langle jj' \rangle }
\mathbf{S}_{i,j}\cdot\mathbf{S}_{i,j'} ,
\label{ham2d}
\end{equation}
where $\mathbf{S}$ are $S=1/2$ spin operators. The isotropic
Heisenberg coupling is assumed both within the chain ($J_{i,j}$ with
$i$ denoting sites in the chain and $j$ denoting different chains) as
well as for the interchain term and $\langle jj' \rangle$ run over
$z_{\perp}$ nearest-neighbor chains. E.g., neutron scattering results 
for pure system \cite{kenzelmann01} 
BaCu$_2$Si$_2$O$_7$ show that the interlayer
coupling is in fact only twice weaker than the intralayer one. Taking
into account also a further non-frustrating diagonal coupling $J_3$
the MFA becomes rather well justified at least on the lowest
nontrivial level. Further more, in the same reference
\cite{kenzelmann01} it has been shown, that for the pure non-random
chain using the MFA with the proper $z_\perp=4$ and
$J_\perp=(1/4)[2|J_x| + 2|J_y| + 4|J_3|]\ll J_z$ 
yields very good estimates for $T_N$ and $m_0$.
Here we used the same notation as in Ref. \onlinecite{kenzelmann01}
with $J_z$ being
intrachain coupling, $J_x$ and $J_y$ interchain couplings and $J_3$
interchain non-frustrating diagonal coupling. 
 We therefore adopt the same $z_\perp$ and use for
comparison to experiments the same $J_\perp$. 
This holds also for doped material,
but with less clear role of disorder on $J_{\perp}$ which we
discuss again in Section \ref{sec:exp}.

Still we expect in analogy to other
quasi-1D spin systems \cite{schulz96,yasuda05} that the main
ordering features should be captured by the MFA for interchain
coupling and by the effective 1D RHC with the staggered field $h_s$ provided
that $J_{\perp} \ll J_i$,
\begin{equation}
H^\textrm{MF}=\sum_{i}J_i\,\mathbf{S}_{i}\cdot\mathbf{S}_{i+1}-h_s\sum_i(-1)^{i}S^{z}_{i} .
\label{hammf}
\end{equation}
Within the MFA the staggered field is given by $h_s=-z_{\perp}J_{\perp} m_s$ with
the staggered magnetization 
$m_{s}=(1/L)\sum_{i} (-1)^i \langle S^{z}_{i}\rangle$ and
$\langle\dots\rangle$ denoting thermal average. We will further on
consider random quenched $J_{i}$ and assume their
distribution to be uncorrelated uniform boxed distribution with
$J-\delta J\le J_{i}\le J+\delta J$ and $\delta J< J$.
For experimental examples more appropriate distribution would be binary
one, but it has been verified \cite{herbrych13} that qualitative
features do not depend essentially on the form of the distribution.
In the following we use units $J=1$ and set
$k_\textrm{B}=\hbar=1$. 

\section{Staggered susceptibility and N\'eel temperature}
\label{sec:ssnt}
Within the MFA for the interchain coupling the instability towards the AFM
ordering and the ordering temperature $T_N$ are determined by the
staggered static susceptibility $\chi_{\pi}$ of a 1D chain and the 
relation \cite{scalapino75,schulz96,joshi03,yusuf05} 
\begin{equation}
z_{\perp}| J_{\perp}| \chi_{\pi}(T_N)=1.
\label{eq_zperpjperp}
\end{equation}
Such a relation is commonly derived within the random phase approximation approach but is
generally coming from the selfconsistency (linear response) relation
at the transition $m_s = \chi_\pi(T_N) h_s$ independent whether the
system is clean \cite{scalapino75,schulz96} or disordered within the
chain \cite{joshi03,yusuf05}. Clearly, it is valid within MFA since
$h_s$ is assumed as the averaged one, while $\chi_\pi(T)$ corresponds
to a macroscopic value (equivalent to disorder averaged one). It is
expected that even in strongly disordered systems the conditions for
Eq.~(\ref{eq_zperpjperp}) are well satisfied for $z_{\perp}
|J_{\perp}| \ll J$. It should however be noted that some quantitative
correction as discussed for clean systems \cite{yasuda05,irkhin00, bocquet02}
($z_\perp \to k z_\perp$ with $k < 1$ due to 
quantum fluctuations) to Eq.~(\ref{eq_zperpjperp}) might be
relevant.

We evaluate $m_s(T)$ and $\chi_\pi(T)$ using the finite-temperature
dynamical density matrix renormalization group (FTD-DMRG) method
\cite{kokalj09,herbsup} on a finite chain with $L$ sites and open
boundary conditions. In the FTD-DMRG method standard
$T=0$ DMRG targeting of ground state density matrix
$\rho^0=|0\rangle \langle0|$ is generalized with finite-$T$ density
matrix $\rho^T=(1/Z)\sum_n |n\rangle \mathrm{e}^{-H/T}\langle
n|$. Next, the reduced density matrix is calculated and then
truncated in the standard DMRG-like manner for basis optimization.
The limitation of the FTD-DMRG method are at low $T$ finite-size
effects, which are rather small due to large accessible system with
DMRG algorithm and which are even further reduced with randomness.

The quenched random $J_i$ are introduced into the DMRG procedure at
the beginning of {\it finite} algorithm. {\it Infinite} algorithm is
preformed for homogeneous system $J_i=J$ and the randomness of $J_i$
is introduced in the first sweep. In this way the preparation of the
basis in the {\it infinite} algorithm is performed just once and for
all realizations of $J_i$-s, while larger number of sweeps (usually
$\sim 5$) is needed to converge the basis within the {\it finite}
algorithm for random $J_i$. After {\it finite} algorithm, magnetization
$\langle S^z_i \rangle$ at desired $T$ is calculated at every site of
the chain within {\it measurements} part of DMRG procedure.
Furthermore, for systems with $\delta J >0$ we employ also
random configuration averaging and typically $N_r=10$ realizations for
finite-$T$ is sufficient due to $\chi_\pi$ being macroscopic quantity
with modest fluctuations between different disorder realizations. For
$T=0$ we use smaller $N_r=5$, since standard DMRG method and larger
systems ($L=800$) can be used. 

 $\chi_\pi$ can be evaluated via dynamical
susceptibility $\chi''(\pi,\omega)$, still we use mostly the
alternative approach by evaluating $m_s$ at finite $T$ and $h_s$, and
then using $\chi_{\pi}(T)=\lim_{h_s\to 0}m_s(T,h_s)/h_s$.  Within this
approach numerical results are more robust or reliable since only
static quantities are calculated and finite size or boundary effects
can be reduced, e.g., by considering only sites close to the middle of
a chain. Still, limit $h_s\to 0$ is hard to reach numerically, but at
finite $T$ small field $h_s\sim 0.01$ suffices.

\begin{figure}[!ht]
\includegraphics[width=0.9\columnwidth]{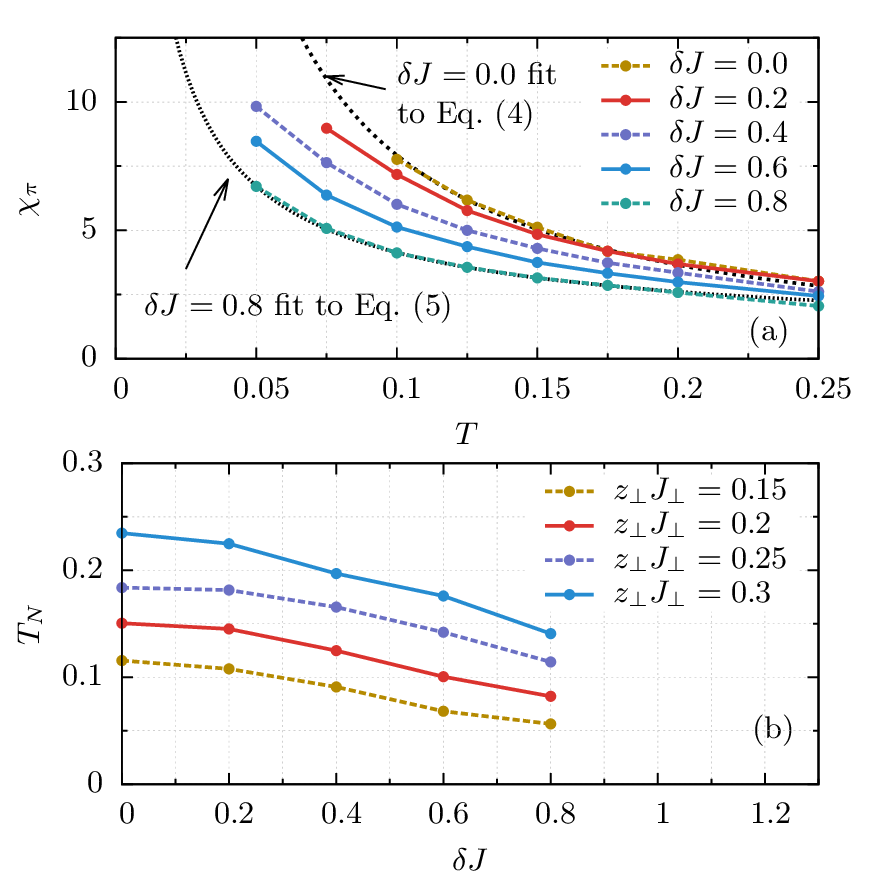}
\caption{(Color online) (a) $T$ dependence of $\chi_\pi$
for various randomness $\delta J$.
Black, dashed line represents RS fit, Eq.~\eqref{chid},
for $\delta J=0.8$.
Shown is also a fit for pure case to Eq.~\eqref{chip}.
(b) Decrease of N\'eel temperature $T_N$ with randomness $\delta J$
for various $z_{\perp}J_{\perp}$. 
Calculated with $L=80$.
}
\label{FT_ms}
\end{figure}

Results for $\chi_\pi$ used to extract $T_N$ with
Eq.~\eqref{eq_zperpjperp} are for several $\delta J$ shown in
Fig.~\ref{FT_ms}a. 
For $\delta J=0$ analytical approaches
\cite{giamarchi89,singh89,affleck89} suggest that for $T \to 0$
\begin{equation}
\chi_{\pi}^p= a\sqrt{\ln\left(b/T\right)}/T\,,\label{chip}
\end{equation}
and also higher order
corrections are discussed \cite{barzykin01}.
Results for random $\delta J\ne0$ 
shown in Fig.~\ref{FT_ms}a clearly indicate that
increasing $\delta J$ reduces $\chi_\pi$ and consequently
leads to a systematic decrease of $T_N$ (for fixed $J_{\perp}$ and
$J$) as shown in Fig.~\ref{FT_ms}b. Fig.~\ref{FT_ms}a also reveals
that $\chi_\pi(T)$ qualitatively changes with increasing disorder.
While for pure case the $T \to 0$ behavior in Eq.~\eqref{chip} is well
 followed, for large $\delta J> 0.5$ we find that
\begin{equation}
\chi_{\pi}^{RS}=c\left[T\ln^2(d/T)\right]^{-1}\,,\label{chid}
\end{equation}
established by RS and with a modified RG approach discussed further
on, fits numerical results better. Since our temperature span is quite
limited ($0.1<T<0.25$ for $\delta J=0$ and $0.05<T<0.25$ for $\delta
J=0.8$) we can not extract precise values of the parameters and even
less comment on the functional forms. However, numerically obtained
staggered susceptibility $\chi_{\pi}(T)$ for random system ($\delta
J=0.8$) is better fitted or described with Eq.~\eqref{chid} than
Eq.~\eqref{chip} and vice versa for the pure case ($\delta J=0$). Note
that in the latter case, quantum Monte Carlo gives
\cite{starykh97,kim98} $a\simeq0.30$--$0.32$ and $b\simeq5.9$--$9.8$,
while for random case this is the first report (see Appendix
\ref{app:tf}) (at least to our knowledge) of the estimated parameter
values.

Experimentally significant $T_{N}/J \lesssim 0.02$
($J_\perp/J\lesssim 0.02$) \cite{yamada01,kenzelmann01}
requires $\chi_\pi \gtrsim 12.5$ (with $z_{\perp}=4$),
which is at present beyond the reach of the FTD-DMRG
method. In order to analyse $T_N$ we chose
modest values of $z_{\perp}J_{\perp}=0.15,\ldots, 0.3$,
presented in Fig.~\ref{FT_ms}b. Still, for the smallest considered
$z_{\perp}J_{\perp}=0.15$ we get reduction of
$T_N$ by a factor of $\sim 2$ for $\delta J=0.8$. This is in
contrast to previous RG study \cite{joshi03} discussed later on,
but in agreement with experimental observations
\cite{thede12,thede14,yamada01}.

\section{Staggered ordered moment}
\label{sec:om}
\begin{figure}[!ht]
\includegraphics[width=0.9\columnwidth]{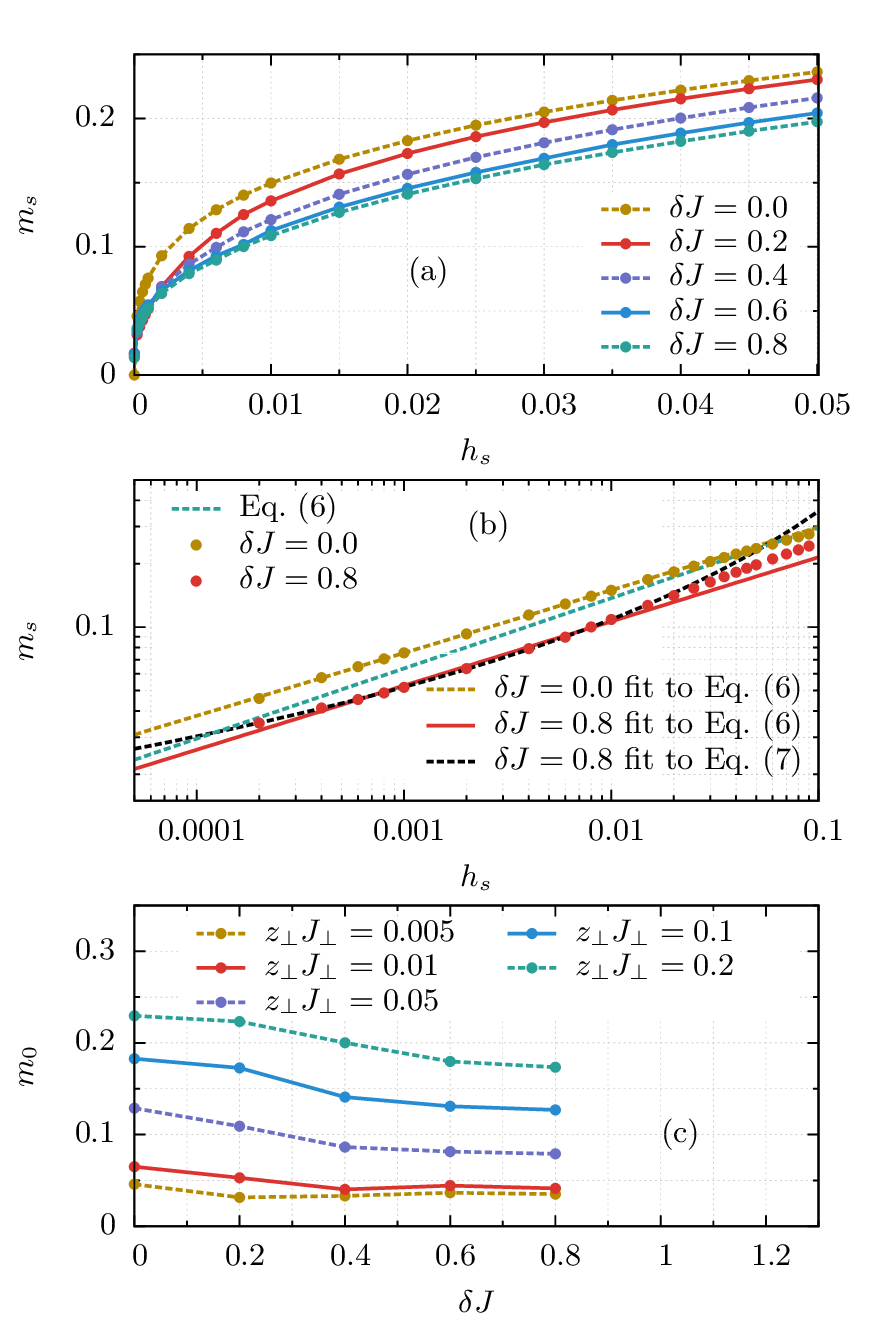}
\caption{(Color online) (a) $T=0$ staggered magnetization $m_s$
vs. $h_s$ for various randomness $\delta J$. (b)
$\textrm{Log}$-$\log$ plot of $m_s$ vs. $h_s$ for $\delta J=0,
0.8$. $m_s(h_s)$ for $\delta J=0$ deviates from prediction in
Eq.~\eqref{m0p} in exponent $g$ and prefactor $r$. The result for
$\delta J=0.8$ shows a RS like behavior given with
Eq.~\eqref{m0rsh}. Fits of parameters for Eq.~\eqref{m0p} or
\eqref{m0rsh} are for regime $0.0001<h<0.01$. (c) Self-consistent
solution for staggered magnetization $m_0$ vs. $\delta J$ for
different $z_{\perp} J_{\perp}$. }
\label{T0_m0}
\end{figure}

In order to determine the $T=0$ average staggered
ordered moment $m_0$ for particular $J_{\perp}$ and disorder $\delta J$
as a solution to MFA self-consistency relation $-h_s/(z_\perp
J_\perp)=m_s(h_s)$,
we first evaluate the g.s. $m_s(h_s)$.
Again finite-size 
effects are largest for the pure case ($\delta J=0$) but in reliable
regime ($h_s>0.0001$) we can make a comparison to the 
analytical result obtained from Ref. \onlinecite{schulz96},
\begin{equation}
m_s^p = r (h_s)^g,
\label{m0p}
\end{equation}
with $r=0.637$ and $g=1/3$. In Fig.~\ref{T0_m0}b we compare Eq.~\eqref{m0p}
to our DMRG results and reveal substantial differences.  Our
$m_s(h_s)$ for $\delta J=0$ shows rather stronger increase with $h_s$,
which cannot be reconciled with Eq.~\eqref{m0p} simply by just
increasing prefactor $r$. Linear dependence shown in Fig.~\ref{T0_m0}b
suggests different exponent ($g\ne 1/3$) or possibly some logarithmic
corrections.

Results in Fig.~\ref{T0_m0}a,b show that
disorder $\delta J$ leads to a decrease of staggered magnetization
$m_s$ in our $h_s$-regime. A possibility of increased
$m_s$ with increased $\delta J$ remains at very low $h_s <0.0001$ as
suggested in Fig.~\ref{T0_m0}b. We investigate and
discuss it later also with the use of RG method. Ordered
moment $m_0$ and its decrease with
$\delta J$ for different values of $z_{\perp}J_{\perp}$ is presented in
Fig.~\ref{T0_m0}c.

\begin{figure}[!ht]
\includegraphics[width=0.9\columnwidth]{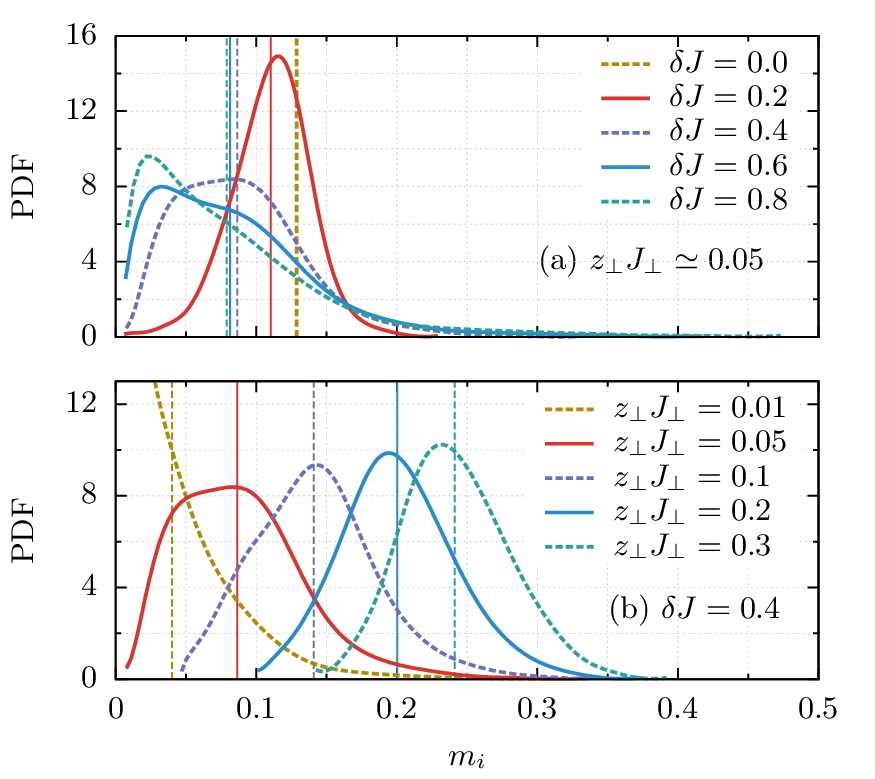}
\caption{(Color online) Probability distribution function of $m_i$ at
$T=0$ for (a) various values of $\delta J$ and fixed
$z_{\perp}J_{\perp} = 0.05$, and (b) for fixed $\delta J=0.4$ and
various $z_{\perp}J_{\perp}$. Thin, vertical lines represent $m_0$
for given $\delta J$ and $J_\perp$.}
\label{T0_sz_pdf}
\end{figure}

A novel feature introduced by disorder is the distribution of local
ordered moments. To avoid the influence of open boundary conditions
we calculate local staggered $m_i=(-1)^i\langle S^z_i\rangle$ from the
middle of the chain modeled with Eq.~\eqref{hammf} and for the MFA
self-consistent fields $h_s$ at particular $z_\perp J_\perp$. Even in
a uniform staggered field $h_s$ moments $m_i$ are found to vary from
site to site and depend on the concrete random configuration $J_i$.
We present the probability distribution function (PDF) in
Fig.~\ref{T0_sz_pdf}a for different randomness $\delta J$ and fixed
$z_{\perp}J_{\perp}=0.05$, while in Fig.~\ref{T0_sz_pdf}b we show it
for fixed $\delta J$ and different $z_{\perp}J_{\perp}$. It is evident
from Fig.~\ref{T0_sz_pdf}a that for large disorder and small
$z_{\perp}J_{\perp}$ the PDF largely deviate from the Gaussian-like
form. Moreover, the relative spread of distribution
$\Delta=\sigma_{m_i}/m_0 $ can become even $\Delta > 1 $.

\section{Real space renormalization group}
\label{sec:rg}
For better understanding and interpretation of above results within
the RS concept we perform similar real space renormalization group
procedure as introduced by Dasgupta and Ma \cite{dasgupta80} and used
also in Ref. \onlinecite{joshi03} and where strongest bonds are eliminated
and reduced effective coupling $J_{\mathrm{eff}}$ is introduced.  We
generalized the procedure for finite $h_s$ and calculation of $m_s$
and give more technical details in the Appendix \ref{app:rs}. We perform
RG procedure numerically on a large system and by carrying it to the
end together with evaluation of staggered magnetization for different
starting staggered fields we obtain $m_s(h_s)$ for $T=0$. A simple RS
argument suggest that in a finite $h_s$ all spins with effective
coupling $J_{\mathrm{eff}}<h_s$ are fully polarized, while the ones
with $J_{\mathrm{eff}}>h_s$ form singlets and contribute only weakly
to the staggered magnetization. Since the portion of spins with
$J_{\mathrm{eff}}<h_s$ in a RS theory is $\propto \ln^{-2}(n/h_s)$
\cite{dasgupta80}, one expects for small $h_s$
\begin{equation}
m_s^{RS}(h_s) \propto \ln^{-2}(n /h_s).
\label{m0rsh}
\end{equation} 
We confirm this RS prediction with our  numerical RG (Appendix \ref{app:rs}),
and $T=0$ DMRG results shown in Fig.~\ref{T0_m0}b 
at low $h_s$, since they deviate from simple power law behavior
of Eq.~\eqref{m0p} (linear in $\log$-$\log$ plot) with a
substantial upward curvature, nicely captured with Eq.~\eqref{m0rsh}.
Our result in Fig.~\ref{T0_m0}b therefore represents one of a few
\cite{hoyos07, hida96,shiroka13} confirmations of the RS phenomenology.

With RG procedure one can make also predictions for finite-$T$
results (see Refs. \onlinecite{dasgupta80,fisher94}), which are 
obtained by
preforming RG steps as long as some Hamiltonian parameter
(e.g. exchange coupling) is larger than $T$, while for the remaining
system with all effective parameters below $T$, a high $T$ result is
used.
In our case
with the system in finite magnetic field $h_s$, these fields do not get
reduced with RG and therefore roughly set the lowest energy
scale. This means that for $T<h_s$ one can perform the RG 
to the end and obtain $T=0$ result for all $T<h_s$. Once $T$ becomes
above $h_s$ all steps with $J<h_s$ cannot be performed
and for this remaining system the
high-$T$ result ($m_s$ roughly linear in $h_s$) should be used.
This leads for $h_s\ll T$ to a random singlet like prediction for
staggered magnetization 
$m_s=h_sc [T\ln^2(d/T))]^{-1}$, and straightforwardly 
for the staggered susceptibility given in Eq.~\eqref{chid}.
Staggered susceptibility has the same functional form as a RS prediction
for uniform susceptibility \cite{fisher94,hirsch80,zheludev07}
$\chi_0(T)$, which can be expected for random system with no
translational symmetry and strongly local correlations. In
Fig.~\ref{FT_ms}a we show that our numerical
calculations with FTD-DMRG give support to this RS prediction.

\section{Comparison with experiment}
\label{sec:exp}
Turning to the experimental realizations of random spin chains, two
systems have been studied so far with magnetic ordering at low $T$,
namely BaCu$_2$(Si$_{1-x}$Ge$_x$)$_2$O$_7$
\cite{yamada01,shiroka11,thede14} and Cu(py)$_2$(Cl$_{1-x}$Br$_x$)$_2$
\cite{thede12}, and for the former a clear evidence of 1D RS physics
has already been detected for $T>T_N$ \cite{shiroka11,herbrych13}. 
Its magnetic
properties can be well described by a simple bimodal distribution of
AFM in-chain exchange constants \cite{shiroka11}, namely $J_i = J_1,
J_2$ with probabilities $x$ and $1-x$, respectively, and by weak
interchain coupling $J_\perp\ll J_i$. Although our treatment assumes a
uniform distribution of the exchange constants, it should be able to
capture general features of BaCu$_2$(Si$_{1-x}$Ge$_x$)$_2$O$_7$,
particularly with Ge concentration $x \sim 0.5$ \cite{herbrych13}.

\begin{figure}[!ht]
\includegraphics[width=0.9\columnwidth]{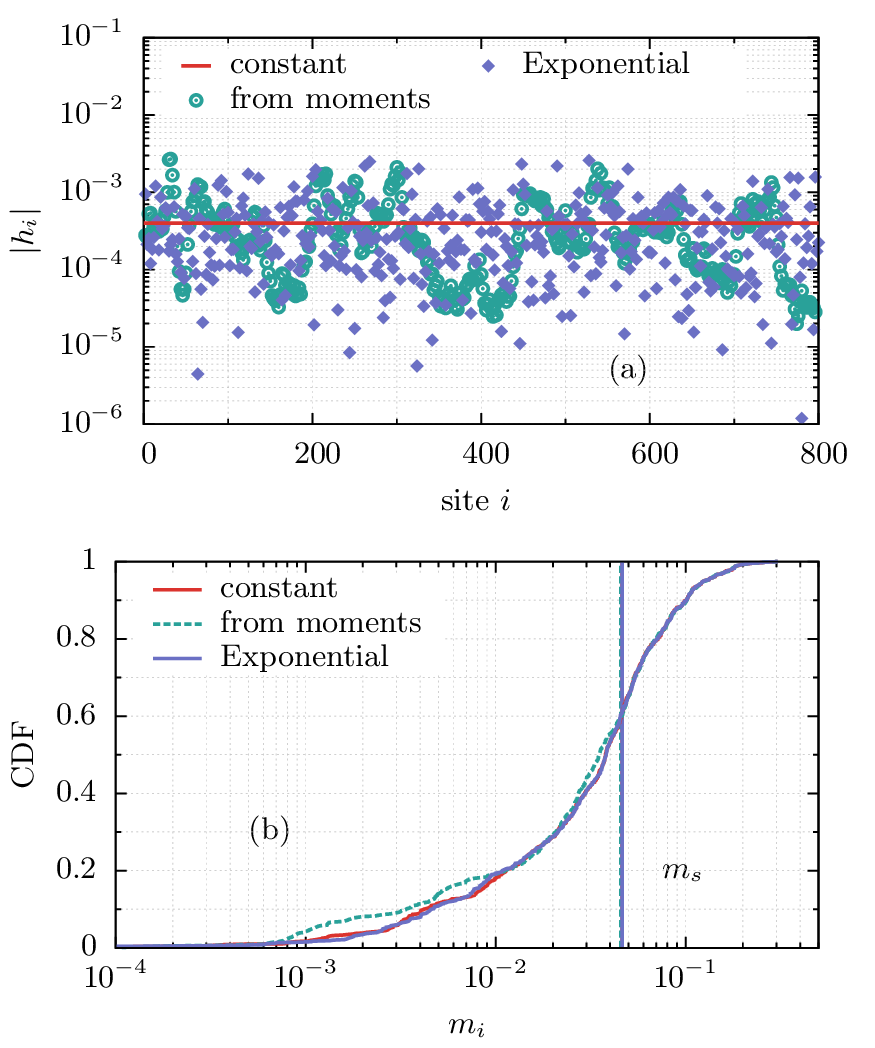}
\caption{(Color online) Dependence of the magnetic moment distribution
(for fixed $J_i$) on the magnetic field used in the mean-field
approximation, as calculated for $L=800$ sites. Panel
(a) shows three distributions of staggered fields with the same average value
$\langle|h_{i}|\rangle=0.0004$; constant one, exponential distribution, and
distribution from numerical simulations (see text for details). 
Panel (b) shows corresponding cumulative distribution function
(CDF) of magnetic moment $m_i$. Vertical lines represent values of average
$m_s$, which are practically the same for all three distributions of $h_i$. }
\label{randfield}
\end{figure}

The experimental data that are most relevant to our calculations are
$\mu$-SR experiments, from which the magnitude of $m_0$ can be
inferred.  In full agreement with our predictions, in both
Cu(py)$_2$(Cl$_{1-x}$Br$_x$)$_2$ \cite{thede12} and
BaCu$_2$(Si$_{1-x}$Ge$_x$)$_2$O$_7$ \cite{thede14}, $m_0$ and the
ordering temperature $T_N$ were found to {\it decrease} with
increasing disorder. This said, the drop in
BaCu$_2$(Si$_{1-x}$Ge$_x$)$_2$O$_7$ appears more abrupt than
predicted. One of the possibility would be that the
strength of $J_\perp$ and even its sign may be locally affected by
disorder as observed in Ref. \onlinecite{yamada01}.
This may also be an indication of MFA limitations and
possibility that a wide distribution with long tails of local moments
or effective local fields used in the MFA (not taken into
account due to used constant averaged field $h_s$), could affect
the results. It is thus
compelling to check, if initial staggered field $h_s$ in Eq. ~\eqref{hammf} should be taken
from adequate distribution of the local moments
$\{h_i\}=-z_{\perp}J_{\perp}\{m_i\}$ and if wide distribution of
initial $h_i$ fields could affect the results in Fig.~\ref{T0_sz_pdf}.
We check this in two ways:
(i) by taking random $h_i$ with exponential distribution,
and (ii) by taking the distribution of
$m_i$ (and thus $h_i$) from another realization of $J_i$.
Fig.~\ref{randfield} depicts comparison between these two methods,
together with constant staggered field (as in Fig.~\ref{T0_sz_pdf})
calculated for $L=800$ and one fixed realization of $J_i$. 
In Fig.~\ref{randfield}a we present three distributions of the
staggered magnetic fields $h_s$ and in Fig.~\ref{randfield}b
corresponding cumulative distribution function (CDF) of magnetic
moment $m_i$. It is clear from the later that the distribution of $m_i$
do not depend strongly on distribution of $h_i$ and even more
importantly for MFA, all considered distributions give very similar
averages of $m_i$. We would like to note that in
Fig. \ref{randfield} we present one of the most critical cases with small
averaged fields with resulting very broad distribution of $m_i$, and
that even in this case the constant fields or fields with very wide
distribution give very similar results for distribution of moments.

The most interesting experimental observation for
BaCu$_2$(Si$_{1-x}$Ge$_x$)$_2$O$_7$ is a drastic broadening of the 
distribution of local static moments in the magnetically ordered
state \cite{thede14}. This behavior is consistent with our predictions
borne in Fig. \ref{T0_sz_pdf}. Unfortunately, making a
quantitative comparison 
beyond a qualitative agreement is not feasible at present, since
$\mu$-SR measures the distribution of local
magnetic fields, not magnetic moments. Due to the presence of several
crystallographic muon sites, the $m_0$ distribution can not be
unambiguously extracted from such experiments.

\section{Conclusions}
\label{sec:diss}
In conclusion, we have shown that at
fixed average $J$ and interchain coupling $J_{\perp}$ the disorder 
$\delta J > 0$ leads to a decrease of N\'eel temperature $T_N$ as well as to
reduced g.s. ordered staggered moment $m_0$, in a very broad range of
$\delta J >0 $ (and regime studied here). This
is due to $\chi_\pi$ being smaller for random system than for a pure
system in a relevant regime (see Fig.~ \ref{FT_ms}a), which is in
contrast with the uniform susceptibility $\chi_0(T \to 0)$ which
approaches constant for pure case but diverges $\propto 1/[T \ln^2
(\tilde \beta /T)]$ for $\delta J > 0$.
 This is analogous to
Eq.~(\ref{chid}) and 
a direct signature of RS 
scenario leading at low $T $ to formation of local singlets and
almost free spins.
The effect of disorder at $q=\pi$ is less dramatic than for $q=0$
since to the leading order (neglecting $\log$ corrections) both
pure and $\delta J>0$ cases reveal $\chi_{\pi} \propto 1/T$. 
However, in a random case $\chi_\pi$ is still larger than
$\chi_0$ (same holds also for structure factor as shown in
Fig. 15 in Ref. \onlinecite{hoyos07}) and the system still tends to AFM order.

Numerical results for $m_s(h_s)$ at $T=0$ in Fig.~\ref{T0_m0}a,b
show that in the regime with larger $h_s$ (e.g., $h_s > 0.0001$ for
$\delta J=0.8$) the average moment $m_s$ (and in turn $m_0$ shown in
Fig.~\ref{T0_m0}c) decreases with increasing $\delta J$. On the other
hand, Eqs.~\eqref{m0p}, \eqref{m0rsh} and
results in Fig.~\ref{T0_m0} suggest a regime of very low
$h_s$ where $m_s$ ($m_0$) could be
increased by $\delta J>0$. This could be relevant only for larger
$\delta J$ and for very small $J_\perp$ ($\lesssim 0.001$ for $\delta
J=0.8$) which would lead to enhanced $T_N$ and $m_0$ with increased
$\delta J$ or in other words, to {\it order by disorder}. Such
behavior
was actually predicted by MFA and RG treatment
\cite{joshi03}, but is contrary to the one mainly discussed here,
as well not found in materials of interest \cite{thede14}.

The most striking effect of the RHC physics and of anomalous RS 
response in the ordered phase is
however the distribution of local moments $m_i$, as manifested by
PDF$(m_i)$ in Fig.~\ref{T0_sz_pdf}. It is evident that the relative
distribution width $\Delta$ increases
with $\delta J$ but even more importantly with decreasing $z_\perp
J_{\perp}$.
This is a clear indication that anomalous width
originates in the RS physics and is not trivially related to initial $\delta
J$. For example, the same or constant $\delta J$ results in increased
relative width of 
distribution ($\Delta$),  if  $z_\perp J_\perp$ is decreased (see
Fig.~\ref{T0_sz_pdf}b). 
It should be noted that for larger $\delta J$ even
$m_i<0$ becomes possible (moments $m_i$ locally opposite to local
fields) \cite{shiroka13}. This means that at small $J_{\perp} \ll J$ and 
strongly reduced $T_N$ the PDF width can become large, i.e. 
$\Delta \sim 1$.

Regarding the experiment, our results of decreasing $m_0$ and ordering
temperature $T_N$ with increasing disorder agree with observations of
the $\mu$-SR experiments on Cu(py)$_2$(Cl$_{1-x}$Br$_x$)$_2$
\cite{thede12} and BaCu$_2$(Si$_{1-x}$Ge$_x$)$_2$O$_7$ \cite{thede14}.
Furthermore, we are able to capture with the microscopic model the
interesting experimental observation of the drastic broadening of the
distribution of local static moments in the magnetically ordered state
of BaCu$_2$(Si$_{1-x}$Ge$_x$)$_2$O$_7$ \cite{thede14}.

\acknowledgements{We acknowledge helpful and inspiring discussions
with M. Thede. We acknowledge the support of the
European Union program (J.H.) FP7-REGPOT-2012-2013-1 no. 316165 and
of the Slovenian Research Agency under program (P.P.) P1-0044 and
under grant (J.K.) Z1-5442.}
\setcounter{equation}{0}
\appendix
\renewcommand{\theequation}{A\arabic{equation}}
\section{Temperature fits}
\label{app:tf}
\begin{table}[h]
\begin{tabular}{c|c|c|c|c|}
\cline{2-5}
                                     & \multicolumn{2}{c|}{Eq.~\eqref{chip}} & \multicolumn{2}{c|}{Eq.~\eqref{chid}} \\ \cline{2-5} 
                                     & $a$             & $b$         & $c$           & $d$           \\ \hline
\multicolumn{1}{|c|}{$\delta J=0$}   & $0.3702$        & $9.8$       & $1898$        & $\to\infty$   \\ \hline
\multicolumn{1}{|c|}{$\delta J=0.8$} & $0.0647$        & $\to\infty$ & $18.55$       & $82.56$       \\ \hline
\end{tabular}
\caption{Values of fitted parameters of Eq.~\eqref{chip} and Eq.~\eqref{chid} to the pure
($\delta J=0$) and random ($\delta J=0.8$) datasets (see Fig.~\ref{FT_ms}).}
\label{fitpar}
\end{table}

\section{RG procedure}
\label{app:rs}
We numerically performed similar renormalization group procedure as
introduced by Dasgupta and Ma \cite{dasgupta80} and modified
it to include the staggered magnetic field $h_s$ and extended it for
calculation of staggered magnetization $m_s$, similarly as done in
Ref. \onlinecite{joshi03}.  In the original procedure the bonds with largest
$J_i$ were eliminated which we replace by subsequent elimination of
bonds with largest $J_i^{xx}$. In the presence of broken rotational
symmetry due to staggered magnetic field $h_s$, $J_i^{xx}$ does not
equal $J_i^{zz}$ at further steps of the elimination process. In the
case of $h_s=0$ the criteria equals to the original one used by
Dasgupata and Ma \cite{dasgupta80} and
$J_i^{xx}=J_i^{zz}$. Justification of $J_i^{xx}$ for elimination
criteria is also that it is the only non-diagonal element of the
Hamiltonian and that for $J_i^{xx}=0$ the ground state is a simple
product state or Ne\'el state, which can be exactly obtained by
arbitrary order of the elimination steps provided that elimination is
performed to the end. For finite-$T$ properties also other energy
scales like $J_i^{zz}$ and $h_i$ are important and need to be
considered.

Once the bond of two sites to eliminate are chosen we integrate them
out by the following procedure. First we calculate eigenstates of the
four site Hamiltonian which consists of two sites to be eliminated
(namely sites 2 and 3) plus two neighboring sites (namely sites 1 and
4). Usually the relevant states which we would like to keep are the
four lowest states and from which we could build effective 
Hamiltonian or the new bond (from site 1 to 4) parameters. However, as the
elimination procedure advances the four lowest states of the four site
Hamiltonian do not necessarily have the character of the ground state
on eliminated bond (sites 2, 3) i.e. they do not all have large
overlap with it and some state with the character of higher lying state
on sites 2 and 3 might become low and among first four low lying
states of the four site Hamiltonian. This does not happen if
$J_{12}^{xx,zz}$ and $J_{34}^{xx,zz}$ are much smaller than
$J_{23}^{xx,zz}$. In such case we choose four eigenstates of the 4
site Hamiltonian with the largest overlap with the ground state on
eliminated two sites (sites 2, 3). These four states span the part of
the relevant low energy Hilbert space that we would like to keep and
are close to the states kept in the second order procedure in
Ref. \onlinecite{dasgupta80}. 

From this four states ($|\psi_i \rangle$ with energy $E_i$,
$i=1,\ldots 4$) we build new effective Hamiltonian for the remaining sites
(sites 1, 4) by first constructing $H_{1234}=\sum_i |\psi_i\rangle E_i
\langle \psi_i |$ and then tracing out the eliminated sites
$H_{14}=\sum_{i_{23}}\langle i_{23}|H_{1234}|i_{23}\rangle$. Here
$|i_{23}\rangle$ are basis states for eliminated sites (sites 2, 3).
New $H_{14}$ is the new Hamiltonian in the basis of remaining sites (1
and 4) and from which one can read new effective parameters like
$J_{14}^{xx}$, $J_{14}^{zz}$, $h_1$, $h_4$ and energy of integrated
out sites $E_{23}$.

Similar procedure can be used for determining the parameters of new
operators that we are interested in. For example, operator
$a_1S_1^z+a_2S_2^z+a_3S_3^z+a_4S_4^z$ is transformed into new operator
$\tilde a_1 S_1^z+\tilde a_4 S_4^z+o_{23}$ after integrating out sites
(2 and 3), while in this case the parameters $\tilde a_1$, $\tilde
a_4$ and $o_{23}$ need to be optimally chosen and small relative error
(typically of $10^{-6}$) can appear by approximating the operator in
the basis for remaining sites (1 and 4) by just three parameters.

In this way one eliminates the two sites, obtains new effective
parameters for the Hamiltonian and operator on the new bond
(connecting site 1 and 4) and can proceed with the new step of RG or
by choosing next two sites to eliminate. The ground state energy and expectation
value of the operator in the
ground state are obtained by preforming the RG to the end (eliminate
all sites) and summing all $E_{23}$ and $o_{23}$ for the energy and
the operator expectation values, respectively.

\bibliography{manumfars}
\end{document}